# On-chip ultra-narrow-linewidth single-mode microlaser on lithium niobate on insulator

Renhong Gao,[1,5] Jianglin Guan,[2,3] Ni Yao,[4] Li Deng,[2,3] Jintian Lin,[1,5,9] Min Wang,[2,3] Lingling Qiao,[1] Zhenhua Wang,[2,3] Youting Liang,[2,3] Yuan Zhou,[1,5] and Ya Cheng[1,2,3,5,6,7,8,10]

[1]State Key Laboratory of High Field Laser Physics and CAS Center for Excellence in Ultra-intense Laser Science, Shanghai Institute of Optics and Fine Mechanics (SIOM), Chinese Academy of Sciences (CAS), Shanghai 201800, China.
[2]XXL—The Extreme Optoelectromechanics Laboratory, School of Physics and Electronic Science, East China Normal University, Shanghai 200241, China.
[3]State Key Laboratory of Precision Spectroscopy, East China Normal University, Shanghai 200062, China.
[4]Research Center for Intelligent Sensing, Zhejiang Lab, Hangzhou 311100, China.
[5]Center of Materials Science and Optoelectronics Engineering, University of Chinese Academy of Sciences, Beijing 100049, China.
[6]Collaborative Innovation Center of Extreme Optics, Shanxi University, Taiyuan 030006, China.
[7]Collaborative Innovation Center of Light Manipulations and Applications, Shandong Normal University, Jinan 250358, China.
[8]Shanghai Research Center for Quantum Sciences, Shanghai 201315, China.
[9]E-mail: jintianlin@siom.ac.cn
[10]E-mail: ya.cheng@siom.ac.cn

**We report an on-chip single mode microlaser with low-threshold fabricated on Erbium doped lithium niobate on insulator (LNOI). The single mode laser emission at 1550.5 nm wavelength is generated in a coupled photonic molecule, which is facilitated by Vernier effect when pumping the photonic molecule at 970 nm. A threshold pump power as low as 200 μW is demonstrated thanks to the high quality factor above $10^6$. Moreover, the linewidth of the microlaser reaches 4 kHz, which is the best result in LNOI microlasers. Such single mode micro-laser lithographically fabricated on chip is highly in demand by photonic community.**

Lithium-niobate-on-insulator (LNOI) has shown strong potential to become an important material platform for photonic integrated circuits (PICs) application because of its outstanding optical performance such as high nonlinear optical coefficient/electro-optic coefficient, broad transparency windows, piezo-electric effect, etc [1-4]. A variety of passive components of PICs such as nonlinear frequency convertors, frequency comb, and quantum light sources have been demonstrated on LNOI wafer over the past a few years [5-11]. Very recently, active components, such as high-speed electro-optic modulators, multi-mode micro-lasers and optical micro-amplifiers have also been reported [12-23]. Among them, the multi-mode microlasers are generated in high-Q Erbium ($Er^{3+}$) doped whispering gallery mode (WGM) microcavities [15-17,20,21,23], and the measured linewidths were on the order of magnitude of 0.001 nm (i.e., ~$1.3×10^8$ Hz) limited by the resolution of spectrometers. The WGM microcavities are designed to have diameters on the level of several tens of micron as a tradeoff between the high quality (Q) factors and small mode volumes V, as the high Q/V ratios are desirable for enabling low threshold lasing actions. The relatively small free spectral range (FSR) of the large-diameter microcavities naturally leads to multi-mode lasing actions. On-chip single mode micro-laser with ultra-narrow linewidth has not been demonstrated on LNOI so far, which, as an attractive light source in LNOI-based PICs, remains to be a critical obstacle.

Here, we experimentally demonstrate a single-mode microlaser at low pump threshold in high-Q $Er^{3+}$ doped microdisk photonic molecules through Vernier effect. The photonic molecule consists of two microdisks of different diameters of 23.1 um ($D_S$) and 29.8 um ($D_L$) separated with a tiny gap of 480 nm. We characterize the single-mode laser by measuring the operation wavelength and linewidth of the laser as well as the threshold pump power. The on-chip sing-mode laser shows an outstanding performance.

The high-Q microdisk photonic molecule was fabricated on $Er^{3+}$ doped thin film lithium niobate, which is bonded on silica buffer layer on lithium niobate substrate. The thickness of the Z-cut thin film is 700 nm, and the doping concentration of $Er^{3+}$ ions is ~1 mol%. The fabrication flow of the microdisk photonic molecule is illustrated in Fig. 1(a), which consists of six major steps. First, a chromium (Cr) layer with a thickness of 600 nm was coated on the thin film wafer by magnetic sputtering. Second, the Cr layer was ablated into the pattern of the coupled dual disks by femtosecond laser micromachining, where the two disks were still physically connected with each other. Third, the lithium niobate thin film exposed to air without the Cr pattern/mask was selectively etched by chemo-mechanical polishing (CMP) [24], thus the pattern of the coupled dual microdisks was transferred from the Cr layer into lithium niobate thin film. Forth, focused ion beam milling (FIB) was carried out with milling depth of 1.3 μm to separate the connected dual microdisks with a gap of 480 nm [25,26]. Fifth, CMP was carried out again for polishing the sidewall of the dual disks, in order to reduce the scattering loss from the sidewall [27]. Finally, the Cr mask was removed completely by chemical wet etching and the fused silica beneath the LNOI photonic molecule was undercut into small pillars by chemical etching in diluted hydrofluoride solution. The scanning electron microscope (SEM) image of the fabricated microdisk photonic molecule is shown in Fig. 1(b).

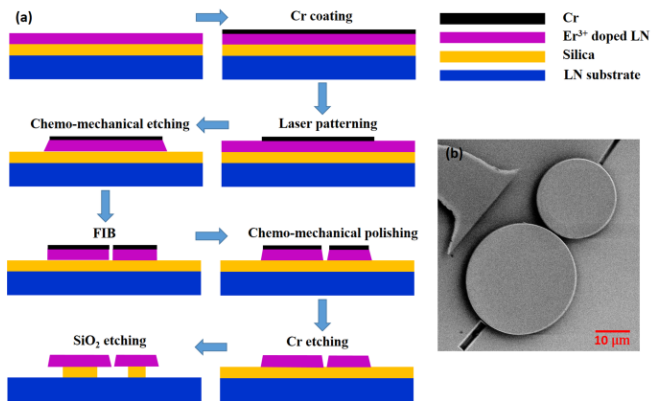

Fig. 1. (a) Illustration of the fabrication flow of the microdisk photonic molecule, and (b) the scanning electron microscope (SEM) image of the fabricated photonic molecule.

To evaluate the Q factor of the photonic molecule, two individual micro-disks of the same diameters as that of the two microdisks in the photonic molecule were fabricated by which the mode structures around 970 nm and 1550 nm wavelengths were measured using a narrow-bandwidth tunable laser with linewidths less than 200 kHz. A fiber taper with a diameter of 1.3 μm was mounted on a 3D piezo stage for achieving critical coupling with the two individual microdisks. WGMs were excited in the two microdisks which gives rise to the transmission spectra as shown by the curves in Fig. 2(a) and (c). The overlaps of the sharp resonant dips in the red and black curves as labeled by blue boxes in the figures indicate that WGM modes at the same wavelengths of 977.7 nm and 1550.5 nm can be excited in both microdisks despite their different diameters. The Q factors of the modes are $1.6\times10^6$ and $1.2\times10^6$ around 970 nm and 1550 nm wavelengths for the small microdisk, respectively, which allow the efficient storage of both pump laser power as well as the generated laser power in the photonic molecule. Consequently, the low pump threshold can be achieved for generating the single mode laser. Importantly, the dips in the two individual microdisks overlap only at 1550.5 nm wavelength in the entire gain spectral range from 1530 nm to 1570 nm of the $Er^{3+}$ ions, allowing for single mode operation of the microdisk laser in the LNOI photonic molecule.

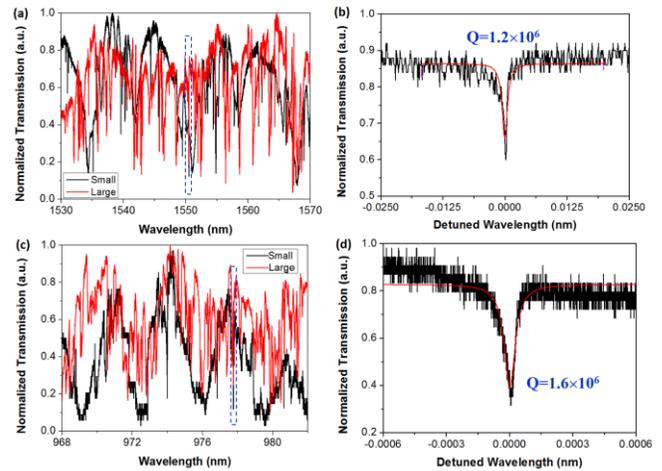

Fig. 2. (a) Transmission spectra of fiber taper coupled with two single microdisks around 1550 nm wavelength (the red curve for the large disk, and the black curve for the small one), (b) The resonant dip of the small microdisk at 977.7 nm wavelength, (c) Transmission spectra of fiber taper coupled with two single microdisks around 970 nm wavelength, and (d) The resonant dip of the small microdisk at 1550. 5 nm wavelength.

Indeed, when we tuned the pump wavelength to 977.7 nm wavelength, single-mode laser was observed at 1550.5 nm wavelength as facilitated by the Vernier effect [28-31], as shown in the spectra in Fig. 3(a). The spectra were collected from the output port of the fiber taper by an optical spectrum analyzer (OSA). The power of the single mode laser increases with the increasing pump power, as shown in Fig. 3(b). The inset in Fig. 3 (b) shows that the pump wave was circulating in both microdisks, although the fiber taper was only coupled with the small disk. We also observed that the laser field was exchanging in the two disks as evidenced by the optical micrograph captured with a long pass filter above the 1100 nm wavelength to remove the pump laser signal, as shown in the inset of Fig. 3(c). These optical micrographs captured at the pump and laser wavelengths indicate that both microdisks are resonant at 977.7 and 1550.5 nm, allowing for generating the single mode laser as a result of the Vernier effect [28-31]. This is owing to the fact that the adjacent resonant wavelength of the photonic molecule is away from $\lambda_0$ 1550. 5 nm with $\Delta\lambda\approx\lambda_0^2/\pi n(D_L-D_S)$=57.1 nm, which is larger than the gain spectrum of ~30 nm. Here, $n$ is the effective index of modes around 1550 nm wavelength, which is about 2.0. The output laser power is plotted as a function of the pump power in Fig. 3(c), showing a linear increase of the output power with the increasing pump power. The threshold of the microlaser was determined as ~200 μW, which is slightly lower than the previously reported threshold pump powers in the multi-mode microdisk lasers on LNOI substrate [15-17].

The linewidth of the microlaser was measured based on laser frequency and phase noise measurement by Michelson interferometer composed of a 3×3 optical fiber coupler [32-35]. The output power of the microlaser was amplified from 300 nW to 10.1 mW by two low-noise Erbium-doped fiber amplifiers (EDFAs), and sent into the Michelson interferometer. Power spectral density (PSD) of differential phase and frequency fluctuation of instantaneous frequency fluctuation of the microlaser and the $\beta$-separation line are shown in Fig. 4(c). Linewidth decreases with the increase of the integration bandwidth. At high frequency domain (>190 kHz), there is only white noise, showing PSD $S_w$ of 110833.27 $Hz^2$/Hz. Thus, the minimum linewidth $\Delta\nu$ (= $\pi S_w$) is 348193 Hz. Background noise induced by EDFAs was calibrated by comparing the measured linewidths of a standard laser (Model: TLB 6728, New Focus Inc.) with and without being amplified by the two EDFAs, which were determined to be 128445.6 Hz and 1474.4 Hz with and without the amplification by the EDFAs, respectively, as shown in

Figs. 4(a) and (b). Thus, the linewidth with the EDFSs was 87.12 times broader than that without the EDFAs. The intrinsic linewidth of the microlaser was measured to be 3996.83 Hz after subtracting the background noise induced by the EDFAs.

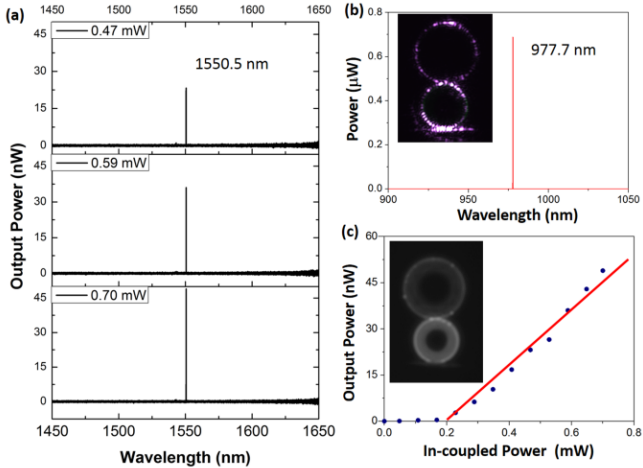

Fig. 3. (a) The increasing single mode lasing signal as the increasing pump power, (b) The spectrum of pump laser, Inset: the optical micrograph of photonic molecule with pump laser, (c) The output power of lasing signal as a function of the pump power, Inset: the optical micrograph of photonic molecule when lasing.

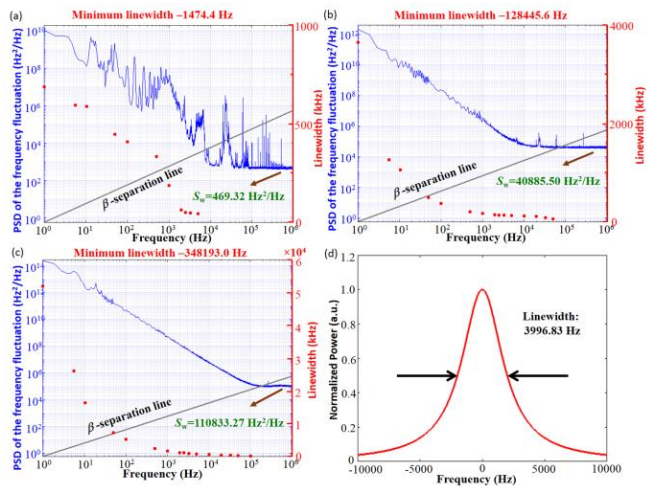

Fig. 4. PSD and linewidth of the standard laser (a) without and (b) with EDFAs. PSD and linewidth of (c) the microlaser with EDFAs. (d) The intrinisic linewidth of the microlaser.

To conclude, a single mode microlaser was demonstrated in coupled microdisk photonic molecule with a pump laser threshold of 200 μW and a linewidth of 4 kHz. The on-chip single-mode microlaser can provide a useful coherent light source for many PIC applications ranging from optical communications and optical computation to precision spectroscopy and metrology, to name only a few.

**Acknowledgments.** We thank Dr. Fei Yang from SIOM for providing support for measuring the linewidth and for instructive discussions.